\documentclass[aps,twocolumn,showpacs,groupedaddress]{revtex4-1}
\usepackage{graphicx}
\usepackage{url}
\usepackage{amssymb,amsmath}
\usepackage{natbib}

\def\apjl{Astrophys. J.}

\begin{document}

\title{Numerical Simulation of the Hydrodynamical Combustion to Strange Quark Matter}

\author{Brian Niebergal$^{1}$, Rachid Ouyed$^{1}$, \& Prashanth Jaikumar$^{2,3}$}

\affiliation{$^1$Department of Physics and Astronomy, University of
  Calgary, 2500 University Drive NW, Calgary, Alberta, T2N 1N4, Canada \\
  $^{2}$Department of Physics \& Astronomy, California State University Long Beach, 1250 Bellflower Blvd., Long Beach, CA 90840 U.S.A\\
  $^{3}$Institute of Mathematical Sciences, C.I.T. Campus, Chennai, TN 600113 India}


\begin{abstract}
We present results from a numerical solution to the burning of neutron matter inside a cold neutron star into stable \textit{u,d,s} quark matter. Our method solves hydrodynamical flow equations in 1D with neutrino emission from weak equilibrating reactions, and strange quark diffusion across the burning front. We also include entropy change due to heat released in forming the stable quark phase. Our numerical results suggest burning front laminar speeds of $0.002-0.04$ times the speed of light, much faster than previous estimates derived using only a reactive-diffusive description. Analytic solutions to hydrodynamical jump conditions with a temperature dependent equation of state agree very well with our numerical findings for fluid velocities. The most important effect of neutrino cooling is that the conversion front stalls at lower density (below $\approx 2$ times saturation density). In a 2-dimensional setting, such rapid speeds and neutrino cooling may allow for a flame wrinkle instability to develop, possibly leading to detonation.

\end{abstract}


\maketitle

\section{Introduction}
\label{sec:intro}
On grounds of asymptotic freedom in Quantum Chromodynamics (QCD), hadronic matter subjected to high densities and/or temperatures will 
deconfine into a quark-gluon plasma. The low-density, high-temperature phase transition happened "in reverse" moments after the Big Bang, and has been fleetingly seen in ultra-relativistic heavy-ion collision experiments (see~\cite{2007Natur.448..302B} for a review). The high-density, low-temperature regime is relevant to compact stars. We assume the Witten hypothesis~\cite{1984PhRvD..30..272W}: bulk strange quark matter (henceforth SQM) is more stable than the nuclear world we live in. The long lifetime of nuclei is reconciled as the improbability of  $\approx A$ weak reactions to occur simultaneously in a nuclear volume containing $A$ nucleons, but SQM can still exist in the form of strangelets or strange quark stars, and co-exist with Neutron stars~\cite{2008arXiv0812.4248B}. Once SQM is nucleated inside a neutron star, how does it grow to form a strange quark star? In this paper, we numerically investigate the issue of combustion of pure neutron matter to \textit{u,d,s} matter using hydrodynamics, taking into account binding energy release and neutrino emission across the burning front 
- going beyond previous treatments of the problem~\cite{1987PhLB..192...71O, 1988PhLB..213..516H,1991NuPhS..24..170O,1994PhLB..326..111C,1994PhRvD..50.6100L,2007ApJ...659.1519D}. This problem is interesting for two main reasons: (i) recent work~\cite{2010arXiv1005.4302H} shows that turbulent effects can increase the front velocity well beyond that expected from laminar flow analysis, entering the distributed regime which is a platform for subsequent detonation and (ii) conversion of a neutron star to a strange quark star has been investigated as an astrophysical model for gamma-ray bursts~\cite{2002ApJ...581L.101L,2002A&A...387..725O,2002NuPhS.113..268B,Haensel:2007tf,2008JPhG...35a4052D}. In this work, we present an improved prescription of the burning front in the laminar flow approximation, and already find speeds as high as $\sim c/100$, where $c$ denotes speed of light. This indicates that unavoidable turbulent effects (such as those discussed in~\cite{2010arXiv1005.4302H}) may well decide the fate of the conversion (deflagaration or detonation). Consider the situation where a compact star's central density has reached that of nuclear deconfinement, and SQM is seeded by one of many possible alternatives~\cite{Alcock:1986}. Recent studies investigated the consequences of such a transition occurring during the core-collapse phase of a supernova \cite{2009PhRvL.102h1101S,2010PhRvD..81j3005D}, or, if nucleation is delayed, in an older neutron star whose central density has increased due to spin-down~\cite{2006ApJ...645L.145S}. 
The conversion scenario we consider is non-premixed combustion~\citep{2000tuco.book.....P} in a cold neutron star,
where SQM (ash) initially grows from a seed by diffusion of strange quarks into 
neutron matter, viewed as a uniform $(udd)$ mixture (fuel). The interface region 
attempts to equilibrate chemically by producing more strange quarks. 
Such a reactive-diffusive setup, assuming a constant-temperature zero-thickness interface, was first explored in~\cite{1987PhLB..192...71O}.
Here, we consider the case for a macroscopically thick interface, evolved with hydrodynamics, paying attention to the temperature gradient and neutrino emission. We find that a self-consistent numerical treatment increases the front velocity by 5-6 orders of magnitude over earlier analytic treatments~\cite{1987PhLB..192...71O,1991NuPhS..24..144H}.  This large difference is mostly due to two assumptions made in previous analytic treatments: (i) considering the fluid and combustion speeds as equivalent,
and (ii)  linearization of the number density difference $n_d-n_s$ in the $d + u\leftrightarrow u + s$ reaction rate.
Combustion inside a fluid involves flame propagation in most cases, requiring a hydrodynamical approach~\cite{2007ApJ...659.1519D}. In addition to the usual conservation equations for the energy-momentum tensor,  
baryon number and electric charge, we also include a diffusion timescale for s-quarks, neutrino emission and entropy evolution due to change in internal energy from converting to SQM. In a typical combustion, local temperature
 increase and subsequent thermal diffusion controls the burning rate.
However, in our situation, the thermal conductivity is small enough~\cite{1993PhRvD..48.2916H,2001ApJ...559L.135U}
that over the simulation time, the temperature gradient across the interface is unchanged. Surprisingly, this temperature variation becomes important through its effect on the pressure, not just reaction rates. We do not include dissipative terms in the hydrodynamical equations.

\section{Hydrodynamics}
\label{sec:hydro}

The 1-D hydrodynamical equations in our case are~\cite{1959flme.book.....L}:
\begin{eqnarray}
\frac{\partial U}{\partial t} &=& -\nabla F\left(U\right) + \mathcal{S}\left(U\right) \ ,
\end{eqnarray}
with variables 
\begin{equation}
\label{eqn:state_vars}
U = \left( \begin{array}{c}
n_s \\
n_s + n_d \\
n_s + n_d + n_u \\
hv \\
s \end{array} \right) \ ,
\end{equation}
and corresponding advective-diffusive terms 
\begin{equation}\label{eqn:adevection}
F\left(U\right) = \left(\begin{array}{c}
vn_s + D\nabla n_s \\
v\left(n_s+n_d\right) \\
v\left(n_s+n_d+n_u\right) \\
hv^2 + P \\
vs \end{array}\right) \ ,
\end{equation}
and source terms
\begin{equation}\label{eqn:sources}
\mathcal{S}\left(U\right) = \left( \begin{array}{c}
-\Gamma_3 + \Gamma_4 + \Gamma_5 \\
-\Gamma_1 + \Gamma_2 - \Gamma_3 + \Gamma_4 \\
0 \\
0 \\
-\frac{1}{T}\sum_{i}\mu_i\frac{dn_i}{dt} \end{array} \right) \ .
\end{equation}
$\Gamma_{1-5}$ are reaction rates for processes in Eqs.(\ref{gamarray})-(13) while index $i$ in the entropy source term ranges over all the particles in the system $i=\left\{u,d,s,e^-,\nu\right\}$. Evolving entropy density $s$, rather than energy density, with a source term describing change in particle species (energy cost of ``assembling'' $(u,d,s)$-matter), allows the binding energy of SQM to be self-consistently taken into account. The enthalpy, $h$, is convenient for fluids that are at relativistic densities. The fluid velocity, $v$, is expressed in units of the speed of light.
To solve this system numerically, we require a constitutive equation (EoS) and the following reactive-diffusive inputs.

{\it Equation of State}: In this work we use the finite-temperature bag model $P = \frac{h}{4} - B$ for the EoS of SQM (neglecting the small electron pressure),
\begin{eqnarray}
\label{eqn:enthalpy}h &=& \frac{19}{9}\pi^2 T^4 + 2T^2\sum_f\mu_f^2 + \frac{1}{\pi^2}\sum_f\mu_f^4 ~\,, \\
\label{eqn:hfromP}s &=& \frac{\partial P}{\partial T}~\,,\\
\label{eqn:nfrommu}n_f &=& \frac{\mu_f^3}{\pi^2} + \mu_f T^2 \ .
\end{eqnarray}
The index $f$ in the above expressions indicates quark flavor (\textit{u,d,s}).
The same EoS is used for both the upstream (unburnt) and downstream (burnt) fluids, with the difference being that the upstream fluid is cold \textit{u,d} matter, since at the point of burning, the neutrons are taken to be already dissolved into a \textit{u,d} fluid (electron pressure is included). We will take up the case of a more complicated EoS, including mixed phases, in subsequent work.

{\it Diffusion and Reactions}: Transport of \textit{d} (fuel) and \textit{s} (ash) quarks through the interface
driven by concentration gradients results in colliding flows of different flavors.
The diffusion coefficient relevant for burning into SQM is~\cite{1993PhRvD..48.2916H}:
\begin{equation}
\label{eqn:diffusion}
D \simeq 10^{-1}\left(\frac{\mu_f}{300~{\rm MeV}}\right)^{2/3} \left(\frac{T}{10~{\rm MeV}}\right)^{-5/3}~{\rm cm^2}/s \ .
\end{equation}

Equilibrium in SQM is established by beta-decay and electron capture reactions,
\begin{eqnarray}
\label{gamarray}
\label{rxn:betadecay_du}d &\rightarrow & u + e^- + \bar{\nu}_e \\
\label{rxn:betadecay_ud}u + e^- &\rightarrow & d + \nu_e \\
\label{rxn:betadecay_su}s &\rightarrow & u + e^- + \bar{\nu}_e \\
\label{rxn:betadecay_us}u + e^- &\rightarrow & s + \nu_e \\
\label{rxn:betadecay_ds}d + u &\leftrightarrow & u + s \ .
\end{eqnarray}

We use rates given by \cite{1997ApJ...481..954A}, which also have equilibrium-seeking terms
for the leptonic processes,
\begin{eqnarray}
\Gamma_1 - \Gamma_2 &=& \frac{34}{5\pi}G_F^2\cos^2\theta_C \\
 &\times& p_{\rm F}\left(d\right) p_F\left(u\right) T^4 \left(\mu_d-\mu_u-\mu_e\right)^2 \\ 
\Gamma_3 - \Gamma_4 &=& \frac{17}{40\pi}G_F^2\sin^2\theta_C\mu_s m_s^2 T^4 \left(\mu_s-\mu_u-\mu_e\right) \\
\label{eqn:d_to_s}\Gamma_5 &=& \frac{16}{5\pi^5}G_F^2\cos^2\theta_C\sin^2\theta_C \\
\nonumber &\times& p_F^2\left(u\right)p_F\left(d\right)p_F^2\left(s\right)\Delta\mu\left[\Delta\mu^2 + \left(4\pi T\right)^2\right] \ .
\end{eqnarray}

where $\Delta\mu=(\mu_d-\mu_s)$, $p_F$ is the quark's Fermi momentum, $G_F$ Fermi's constant and $\theta_C$ the Cabbibo angle. For process (\ref{rxn:betadecay_ds}) to proceed in a given region, a minimum number of \textit{s} quarks must be present. While this number should depend on factors 
such as the strangelet mass and surface tension, here we simply make sure to avoid unphysical effects, such as superluminous diffusion speeds~\cite{1959flme.book.....L},
by imposing a smooth cut-off on the \textit{s} quark Fermi momentum ($p_{F_s}$=$\sqrt{\mu_s^2-m_s^2}\gtrsim0.1$ MeV) for reaction (\ref{rxn:betadecay_ds}) to proceed (this is analogous to the activation temperature, in Arrhenius-type reactions, typically used in modeling heat-diffusion driven combustion).

{\it Neutrino emission}: Neutrinos are emitted copiously from the location of the interface, where the leptonic weak reaction rates from chemical equilibration are highest. At these temperatures (tens of MeV) and densities ($\rho\sim 10^{15}$g/cc), neutrino mean free paths $\lambda$ are 
on the order of $100$ cm~\cite{1982AnPhy.141....1I}. 
Accurate neutrino transport requires solving the Boltzmann equation, which in our setup introduces additional stiffness in the flow equations. A simpler estimate capturing the essential physics in 1-D is to introduce an exponential cut-off on the neutrino emissivity as follows

\begin{equation}\label{eqn:neutrino_cooling}
\varepsilon = \left(\varepsilon_{q\beta} + \varepsilon_{qs}\right)\times e^{-\left(x_{\rm I}-x\right)/\lambda} \ .
\end{equation}

where $\varepsilon_{q\beta}$ and $\varepsilon_{qs}$ denote the non-equilibrium neutrino emission rate for reactions (\ref{rxn:betadecay_du}) and (\ref{rxn:betadecay_su}) respectively~\cite{1997ApJ...481..954A}, $x$ is the position of the emitting region and $x_{\rm I}$ is the position of the front of the burning interface. Effectively, for a given emitting region at $x$, if the the distance to the interface $x_{\rm I}-x$ is more than the mean free path $\lambda$, then produced neutrinos are trapped, otherwise they escape. Since $\lambda\sim 100$cm, neutrinos produced near or in the interface and directed outwards essentially free stream. The matter ahead of the burning interface is 
cool, while SQM behind it is hot and produces many neutrinos. The small mean free path then implies that neutrino cooling does not significantly alter the 
temperature of equilibrated SQM on the timescale of the
simulation, but has an important effect on the diffusion of strange quarks across the interface, and hence the speed of the burning front (Fig.~\ref{fig:v_jumps}). 

\section{Numerical Simulations \& Results} 
\label{sec:results}
The variables in the equations of hydrodynamical combustion~(Eq.~\ref{eqn:state_vars}) are solved for numerically using a fourth-order Runge-Kutta scheme. Spatially, a third-order upwinded advection, flux-limited, finite-volume approach is used \cite{1998cmaf.conf....1L}. The diffusion and pressure gradient terms are second-order, not upwinded, and treated separately from the advection terms (ie. not flux-limited). A large pressure wave is created from the initial state, even though it initially satisfies pressure equilibrium. This is typical for combustion problems, where the (unburnt) fluid in front of the interface is set in motion~(\cite{1959flme.book.....L}, pg. 487).
However, this wave is transient and quickly flows past
the burning region, increasing the speed of the interface without impacting
its long-term evolution (for animations, visit http://www.capca.ucalgary.ca)
An acceptable grid spacing is $\Delta x = 0.05$, resulting in a limited timestep of $\Delta t/\Delta x < 0.3$ from the advection terms, 
and $\Delta t/\left(\Delta x\right)^2 < 1/D$
from the diffusion terms. Leaving more detailed description of numerical aspects to a subsequent article, we discuss here our main physical results.

\begin{figure}[ht!]
\includegraphics[width=0.5\textwidth]{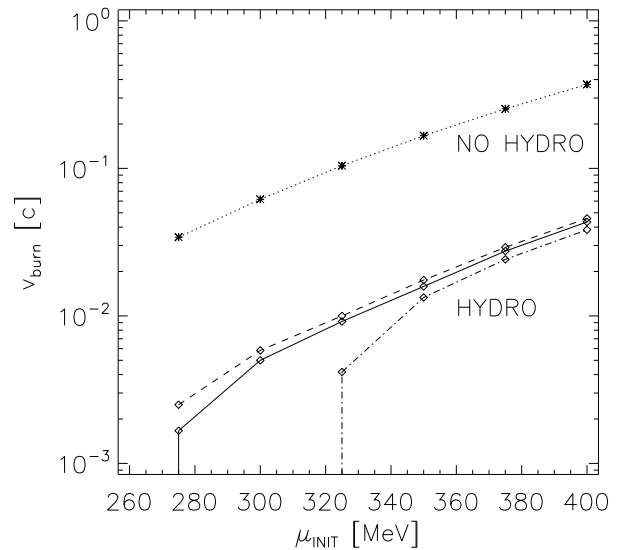}
\caption{\label{fig:interface_speed} Steady-state burning interface speeds $v_{\rm burn}$ for simulations with various initial densities (quark chemical potential) $\mu_{\rm INIT}$.  
The three hydrodynamic cases (HYDRO) are with no neutrino cooling (dashed line), neutrino cooling from Eq.\ref{eqn:neutrino_cooling} (solid line), and enhanced neutrino cooling (dash-dotted line).  The dotted line indicates simulations without hydrodynamics (ie. fluid velocities zero everywhere).
$v_{\rm burn}$ increases with larger densities, since more fuel is present, and decreases with larger cooling rates. As cooling becomes more effective, the hydrodynamic jump conditions (Eq.~\ref{eqn:hydrojumps}) are satisfied by increasingly opposing the advance of the interface, which consequently stalls at progressively higher densities.}
\end{figure}

\begin{figure}[ht!]
\includegraphics[width=0.5\textwidth]{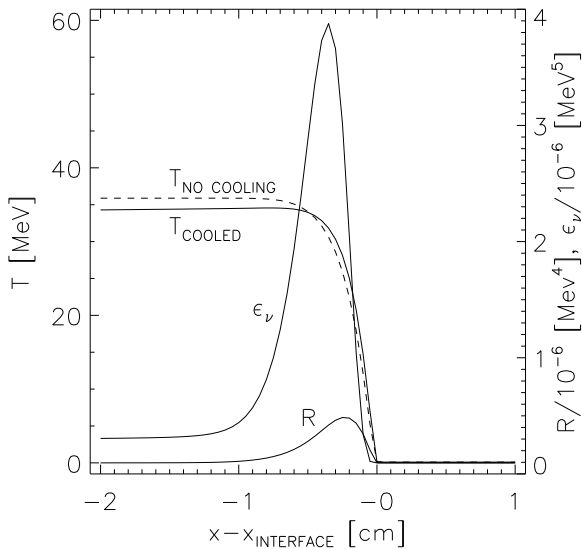}
\caption{\label{fig:rxn-dfsn_profiles2} A snapshot during the simulation of the temperature ($T$), reaction rate ($R$), and neutrino emissivity, $\epsilon_{\nu}$, throughout the burning interface. The temperature is shown with (solid line) and without (dashed line) neutrino cooling effects, where the difference between the two is  the variable $C = T - T_{\rm cooled}$ that serves as the measure of cooling.}
\end{figure}

(i) Effects of hydrodynamics: In Fig.~\ref{fig:interface_speed} the interface speed, with and without the effects of hydrodynamics, is plotted for various initial conversion densities. In the former case, typical speeds for the burning interface were found to be between $0.002c$ and $0.04c$ for initial baryon densities ranging from $1.7 n_0$ to $5.3 n_0$, where $n_0$ is nuclear saturation density. These burning speeds are much higher than previous estimates~\cite{1987PhLB..192...71O,1991NuPhS..24..144H}. The reasons
come from the $T$ and $\mu_s$ variations across the finite-width interface.
Just after contamination, at small values of $\mu_s$, the reactions producing \textit{s} quarks are dominated by the $\Delta\mu^3$ factor in Eqn.(\ref{eqn:d_to_s}). Further behind the interface, \textit{s} quark production becomes increasingly dependent on the temperature term. 
This increases the reaction rate as expected~\cite{1993PhRvD..47..325M}, resulting in a faster speed of the burning interface.
Including hydrodynamics in the reactive-diffusive simulations 
creates different fluid velocities on either side of the interface,
which ends up effectively opposing the interface's progression (discussed below). Typical widths of the interface (see Fig.~\ref{fig:rxn-dfsn_profiles2}) were found to be $\sim 1$ cm when hydrodynamics is included and $\sim 10$ cm for a purely reaction-diffusion system. 

\begin{figure}[ht!]
\includegraphics[width=0.5\textwidth]{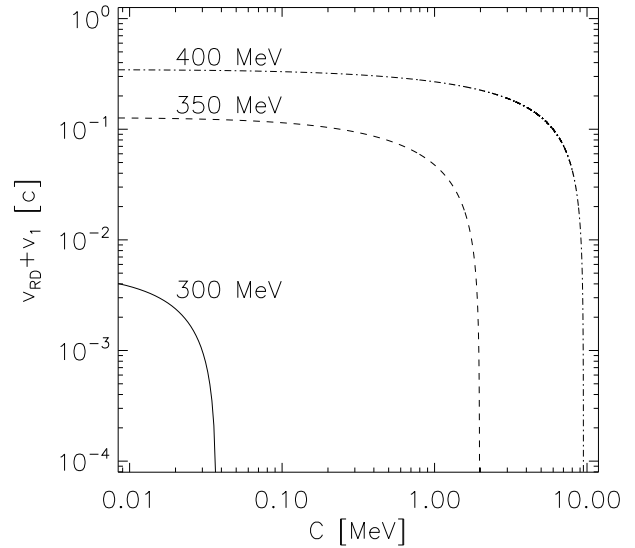}
\caption{\label{fig:v_jumps} Velocity of the burning interface, purely from the reaction-diffusion process ($v_{\rm RD}$; ie. no hydrodynamics)
plus the upstream fluid velocity ($v_1$), versus cooling.  
The upstream velocity is calculated analytically from the jump conditions (cf. Apx. \ref{sec:jump_conditions}), and neutrino cooling is represented by the difference between non-cooled and cooled downstream temperatures 
$C = T - T_{\rm cooled}$. Values shown are the initial densities. The interface halts after a critical amount of energy is removed by cooling. }
\end{figure}

(ii) Effects of $\nu$-cooling: Neutrino emission (deleptonization) causes a decrease in pressure for the burnt fluid. 
The resulting pressure gradient forces fluid velocities to 
become increasingly negative (in the reference frame comoving with the burning interface), causing advection to 
oppose the progression of the burning interface~\footnote{An analytic treatment using the jump conditions as described in the appendix confirms this effect.}. Since cooling rates may have uncertainties, we parameterize the efficacy of neutrino cooling by $C = T - T_{\rm cooled}$, where $T_{\rm cooled}$ and $T$ are the downstream (burnt) temperatures with and without cooling. 
As shown by the two temperature profiles in Fig.~\ref{fig:rxn-dfsn_profiles2},
Even a modest drop in temperature can decrease the pressure enough to enter an advection dominated regime, where the upstream fluid velocity ($v_1$)
advects the interface backwards faster than it can progress due to reactions and diffusion ($\left|v_1\right| > v_{\rm RD}$). In such a case the interface halts, 
as seen in Fig.~\ref{fig:v_jumps}, as soon as $v_{\rm RD} + v_1 < 0$.
This is because as the interface stops diffusing into the fuel, the reactions are no longer proceeding. Since neutrino production drops as a consequence, energy is no longer being removed from the burning region and the system reaches a situation where diffusion and advection are in balance.

\section{Discussion \& Conclusions}
\label{sec:conclusions}
We have performed 1-D numerical simulations of the burning of neutron matter to strange quark matter (SQM) with consistent treatment of reactions, diffusion, and hydrodynamics. By modeling a region of SQM surrounded by \textit{u,d} matter, interpreted as neutron matter above nuclear densities, we find typical speeds of the burning process to be between $0.002c$ and $0.04c$ and interface widths of $\sim 1$ cm. These speeds are noticeably higher than estimates found in previous works, for eg., \cite{1991NuPhS..24..170O,1993PhRvD..48.2916H}. In this work we have addressed the importance of evolving temperature self-consistently from the binding energy release during conversion to SQM by incorporating this with the entropy evolution equation. We have also shown how 
neutrino cooling can halt the burning interface by decreasing pressure support against advective forces. Hence, the importance of neutrinos cannot be overstated and must be addressed more thoroughly in future work.
An equally important focus for future work is a two-dimensional treatment.
While cooling can only halt the interface in one dimension, in two or more dimensions we propose that a new type of instability would develop, caused by regions along the burning interface halting due to cooling, at which point unburnt material starts to flow backwards onto the interface (as inferred from the jump conditions, Apx. \ref{sec:jump_conditions}), whereas regions not halted by cooling will have unburnt material flowing away from the interface. The result is a wrinkled interface, with shearing between the unburnt fluids of halted and non-halted regions. A wrinkled interface increases the diffusion rate, and causes an overall increase in the burning interface's speed ($v_{\rm burn}$). However, the wrinkling is also subject to stabilization by diffusion - in the dimension along the interface concave regions are accelerated while convex regions are decelerated.  
The interplay between stabilization and the wrinkling instability can result in three scenarios: either (i) stabilization is too strong causing 
$v_{\rm burn}$ to remain small and the entire interface halts, 
or (ii) stabilization is moderate and the interface progresses outwards as a combustion, or (iii) stabilization is weak and $v_{\rm burn}$ increases without bound, presumably resulting in a detonation. Validating these options would require high-resolution multi-dimensional simulations, which we leave for future work.

\begin{acknowledgments}
  This research is supported by grants from the Natural Science and
  Engineering Research Council of Canada (NSERC) and Alberta Innovates (iCore).
  P.J. acknowledges support from start-up funds at California State University Long Beach.
\end{acknowledgments}

\bibliography{prc_v2}

\appendix
\section{Verification of Numerical Solutions}
\label{sec:jump_conditions}

We performed numerical tests separately for the diffusive, reactive, and hydrodynamic parts of the code. With only diffusion, we run the usual tests of a diffusing initially gaussian profile and find a relative error in the gaussian width that is smaller than the resolution
at all times. The reactive part of the code confirmed analytically estimated
timescales to achieve weak equilibrium~\cite{1987PhLB..192...71O, 1993PhRvD..47..325M, 1997ApJ...481..954A}. For hydrodynamics, we solved jump conditions in the frame of the burning interface, including the temperature increase due to the release of binding energy. From Eq.(\ref{eqn:adevection}), 
\begin{eqnarray}\label{eqn:hydrojumps}
\left(\mu_{u,1}^3 + \mu_{d,1}^3\right)v_1 &=& \left(\mu_{u,2}^3 + \mu_{d,2}^3 + \mu_{s,2}^3\right)v_2 \\
h_1 v_1^2 + P_1 &=& h_2 v_2^2 + P_2 \ .
\end{eqnarray}
The subscripts $1$ and $2$ indicate upstream (unburnt) and downstream (burnt) fluids respectively. Pressure is given in terms of enthalpy ((Eq.~\ref{eqn:enthalpy}). The above expressions are solved analytically, 
and upstream and downstream velocities agree to better than 2\% with those
found numerically. We do not include the jump condition from the entropy equation, since the reaction term introduces a non-linear component, so the temperature increase due to release of binding energy is not found analytically. Instead, we used computed values from simulations without hydrodynamics.

\end{document}